\documentclass{article}

\usepackage{arxiv}

\usepackage[utf8]{inputenc} 
\usepackage[T1]{fontenc}    
\usepackage{hyperref}       
\usepackage{url}            
\usepackage{booktabs}       
\usepackage{amsfonts}       
\usepackage{nicefrac}       
\usepackage{microtype}      
\usepackage{amsmath}
\usepackage{cleveref}       
\usepackage{upgreek}
\usepackage{bm}
\usepackage{diagbox}
\usepackage{graphicx}
\usepackage[numbers]{natbib}
\usepackage{doi}
\usepackage[version=4]{mhchem}

\title{On-Chip Shaping of Surface Acoustic Waves via Continuous Diffractive Modulation}

\date{}

\author{ {Yuanpeng Ma, Yang Zhao, Shangyang Chen, Dachuan Sang, Sicheng Wang, Dong Zhang\thanks{dzhang@nju.edu.cn}, and Xiasheng Guo\thanks{guoxs@nju.edu.cn}} \\
	Key Laboratory of Modern Acoustics (MOE), School of Physics, \\
	Collaborative Innovation Center of Advanced Microstructures, Nanjing University,\\
	Nanjing 210093, China\\
}

\hypersetup{
pdftitle={On-Chip Shaping of Surface Acoustic Waves via Continuous Diffractive Modulation},
pdfsubject={Piezoelectric acoustics, On-chip field shaping, Diffractive wave manipulation},
pdfauthor={Yuanpeng Ma, Dong Zhang and Xiasheng Guo},
pdfkeywords={Surface acoustic waves, Diffractive wave manipulation, Inverse design, Piezoelectric acoustics, Acoustic field shaping},
}

\begin{document}
\maketitle

\begin{abstract}
Precise shaping of surface acoustic waves~(SAWs) on anisotropic piezoelectric substrates is complicated by elastic anisotropy and electromechanical coupling. Here we introduce a continuous diffractive acoustic lens~(CDAL), in which weak phase-velocity perturbations produced by subwavelength metallization, together with diffraction, drive lateral field redistribution. An anisotropic thin-element propagation model enables efficient inverse design of the CDAL contour. Laser Doppler vibrometry verifies prescribed lateral field profiles and multichannel beam splitting with unequal widths. By combining continuous diffractive modulation with discrete source-phase encoding, we further shape the lateral field into a Bessel profile, demonstrating joint amplitude-and-phase control. This work provides a high-precision and fabrication-compatible mechanism for on-chip SAW field modulation.
\end{abstract}

Controlling the propagation of waves on a chip is a challenge with broad implications. Surface acoustic waves~(SAWs), which confine elastic energy near a solid surface, have been widely exploited for sensing~\cite{gao2018Dual, liang2023Temperature}, acoustofluidics~\cite{collins2018SelfAligned, yang2022Harmonica}, optomechanics~\cite{balram2016Coherent, iyer2024Coherent}, and quantum acoustics~\cite{schuetz2015Universal, satzinger2018Quantum}. As these applications move beyond wave generation and propagation toward elaborate spatial manipulation, precise shaping of SAW fields becomes increasingly important. On piezoelectric substrates, however, SAW propagation is jointly governed by elastic anisotropy~\cite{weser2020Complexity}, electromechanical coupling~\cite{farnell1972Elastic} and diffraction~\cite{ma2025Minimizing}, making flexible on-chip field shaping challenging.\par

Currently, SAW modulation primarily relies on either source engineering or artificial structures on the propagation path. Source engineering encodes the initial field through the geometry~\cite{ma2024Generating,li2025Onchip} or independent actuation~\cite{ma2026Multidimensional} of interdigital-transducer~(IDT) elements, thereby enabling focused, self-bending, and other special SAW beams. Such approaches effectively encode the initial beam phase, yet the resulting SAW beam remains constrained by the transducer geometry. Local electrode  modification also generally affects electroacoustic conversion efficiency, beam amplitude, and phase simultaneously. Alternatively, phononic crystals~\cite{gao2023Onchip, zhang2024Gigahertz} and metasurfaces~\cite{wang2025Experimental} manipulate propagating SAWs through local resonance or scattering, enabling guiding, splitting, and deflection. These approaches substantially expand propagation-domain modulation but typically rely on pronounced local responses. To date, SAW field shaping still lacks a mechanism that combines a clear physical picture, computational efficiency, and fabrication simplicity.\par

Here we introduce a fundamentally different principle: instead of strong local modification, we exploit the gradual accumulation of weak perturbations coupled with intrinsic diffraction. Based on this principle, we propose a continuous diffractive acoustic lens~(CDAL). The CDAL comprises a subwavelength-thick metal film with a designed two-dimensional contour, which slightly reduces the SAW phase velocity in the lens region~\cite{auld1973Acoustic}, and the induced phase difference accumulates as the SAW propagates across the film. This distributed phase accumulation continuously interacts with intrinsic diffraction, progressively reconstructing the SAW field.\par

This mechanism differs from a phase element~\cite{melde2016Holograms} that is thin along the propagation direction, which imposes a local phase delay within that thin region, and from a conventional gradient-index medium~\cite{lin2009Gradientindex}, whose material parameters vary continuously in space. Locally, a CDAL supports only two propagation states, free and metallized surfaces, yet this binary modulation varies continuously along the propagation direction. Consequently, the output field is not governed by a local transmission function within any single thin region, but by the entire propagation process. From this perspective, the CDAL resembles a multilayer diffractive cascade~\cite{lin2018Alloptical} in optics, except that the modulation-propagation sequence is continuous rather than discrete. The CDAL can therefore be viewed as a propagation-continuous diffractive cascade, in which diffraction, typically associated with beam spreading, is instead exploited as a controllable mechanism for lateral SAW field redistribution.\par

To realize this mechanism, we develop an anisotropic thin-element propagation model and an efficient inverse-design framework for the CDAL contour. Experiments demonstrate accurate reconstruction of target lateral field profiles and multichannel beam splitting with individually defined widths. By further combining continuous diffractive modulation with discrete source-phase encoding, we generate non-diffracting SAW fields~(Bessel beams) that satisfy prescribed amplitude and piecewise phase distributions. These results establish the CDAL as a high-precision, fabrication-compatible approach to on-chip SAW field shaping.\par

\begin{figure}[h!]
 \centering
 \includegraphics[width=8.6cm]{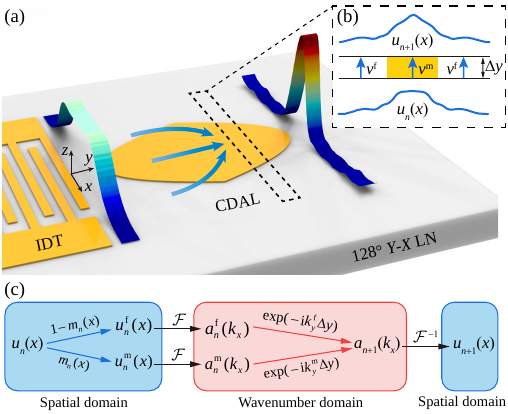}
 \caption{(a)~Schematic of on-chip SAW field shaping via the CDAL. (b)~Local phase shifts and field evolution over a thin modulation element. (c)~Schematic representation of the TED method for calculating the SAW field evolution.}
 \label{fig1}
\end{figure}

Fig.~\ref{fig1}(a) illustrates the basic configuration of the CDAL. A patterned metal film deposited on the piezoelectric substrate defines a weakly perturbed region on the SAW propagation path. As shown in Fig.~\ref{fig1}(b), the slightly different phase velocities on the free and metallized surfaces produce a local phase delay, which accumulates during propagation and modifies the subsequent diffractive evolution. Their combined action progressively reshapes the lateral field into the target profile.\par

For a metal film of thickness $d$ much smaller than the SAW wavelength $\lambda_\mathrm{s}$, its effect can be treated as a perturbation to the free surface~\cite{auld1973Acoustic}. The metallized surface exhibits a slightly lower phase velocity, while the local eigenmode changes only weakly~\cite{yamada1990Scattering}. For \ce{Au} metallization on the $128^{\circ} $\textit{Y}-\textit{X} \ce{LiNbO3} (LN) substrate, the phase velocity decreases by approximately 2.7\%--10\% for $0 \le d/\lambda_\mathrm{s} \le 0.01$. In this weak-perturbation regime, forward propagation dominates. Since the Rayleigh-type SAW is the dominant mode on this substrate~\cite{shibayama1976Optimum}, we model the SAW field using a forward scalar diffraction model~\cite{fagerholm1997SAW}, with the complex out-of-plane displacement $u(x,y) = u_z(x,y,z = 0)$. Anisotropy is incorporated in the propagation analysis through the slowness relation $k_y(k_x)$.\par

The incident field at the CDAL input is determined by the straight IDT geometry and can be calculated using an equivalent line-source model and the angular-spectrum method~\cite{ma2025Minimizing}. Within the CDAL, propagation is governed by the binary distribution of free and metallized surfaces and their respective slowness relations, $k_y^\mathrm{f}(k_x)$ and $k_y^\mathrm{m}(k_x)$. We therefore employ the thin-element decomposition~(TED) method~\cite{fagerholm1997SAW}, discretizing the CDAL along $y$ into thin modulation elements of thickness $\Delta y$.\par  

For the $n$th element shown in Fig.~\ref{fig1}(b), we define a binary mask $m_n(x)$, with $m_n(x) = 1$ on the metallized surface and $m_n(x) = 0$ on the free surface. As illustrated in Fig.~\ref{fig1}(c), the incident field $u_{n}(x)$ is decomposed into the free-surface component $u^\mathrm{f}_{n}(x) = [1-m_n(x)]u_n(x)$ and metallized component $u^\mathrm{m}_{n}(x) = m_n(x)u_n(x)$. After transformation to wavenumber domain, the two components experience different phase shifts and are recombined to give~\cite{fagerholm1997SAW}
\begin{equation}
u_{n+1}(x)=\mathcal{F}^{-1}\Big[
\mathcal{F}\left\{u^{\mathrm{f}}_n(x)\right\}
e^{-\mathrm{i} k_y^{\mathrm{f}}\Delta y} +
\mathcal{F}\left\{u^{\mathrm{m}}_n(x)\right\}
e^{-\mathrm{i} k_y^{\mathrm{m}}\Delta y}
\Big].
\label{eq_TED}
\end{equation}

Eq.~(\ref{eq_TED}) captures the core mechanism of continuous diffractive modulation. Each thin modulation element introduces only a weak local phase perturbation, yet the modulation is distributed throughout the device rather than imposed within a single thin region. As the wave propagates, the free and metallized components repeatedly undergo decomposition, propagation, and recombination. Phase changes generated in one element are converted into amplitude variations by subsequent diffraction, while the evolving field continues to experience new local modulations. The continuously varying CDAL boundary therefore allows small local phase shifts to accumulate into substantial nonlocal lateral redistribution. The output field is thus governed by the entire CDAL contour rather than by a local transmission function.\par

Iterating Eq.~(\ref{eq_TED}) along $y$ across the CDAL yields an efficient forward model from a given contour to the output field. Since the spatial domain and wavenumber domain transformations are implemented via fast Fourier transforms, the model is well suited to iterative inverse design. We parameterize the upper and lower CDAL boundaries using smooth cubic B-spline curves and optimize their geometries to match a target output field. Details of the parameterization and optimization procedure are given in the Supplemental Material.\par

Owing to the nondispersive nature of SAWs, the free-surface phase velocity $v^\mathrm{f}$ is independent of the SAW wavelength $\lambda_\mathrm{s}$. Within the weak-perturbation regime considered here, propagation on the metallized surface is determined by the normalized thickness $d/\lambda_\mathrm{s}$ and the corresponding slowness relation $k_y^\mathrm{m}(k_x)$.\cite{farnell1972Elastic} The CDAL modulation mechanism and design framework are therefore not limited to a specific frequency or wavelength and can be extended to SAW devices operating at different scales, provided that $d/\lambda_\mathrm{s}$ remains within the perturbative regime. \par

As a representative implementation, we used straight IDTs with a finger period of $\lambda = 200~\mathrm{\upmu m}$ as sources in this work, corresponding to a SAW wavelength of $\lambda_\mathrm{s} \simeq \lambda$. The IDTs and CDALs were co-fabricated on a $128^{\circ} $\textit{Y}-\textit{X} \ce{LiNbO3} substrate with an \ce{Au}/\ce{Cr} metallization of $195/5~\mathrm{nm}$, corresponding to $d/\lambda_\mathrm{s} \simeq 0.001$. Under these conditions, the SAW phase velocity on the metallized surface is approximately 3.5\% lower than that on the free surface.\par

\begin{figure*}[bp]
\centering
  \includegraphics[width=17.8cm]{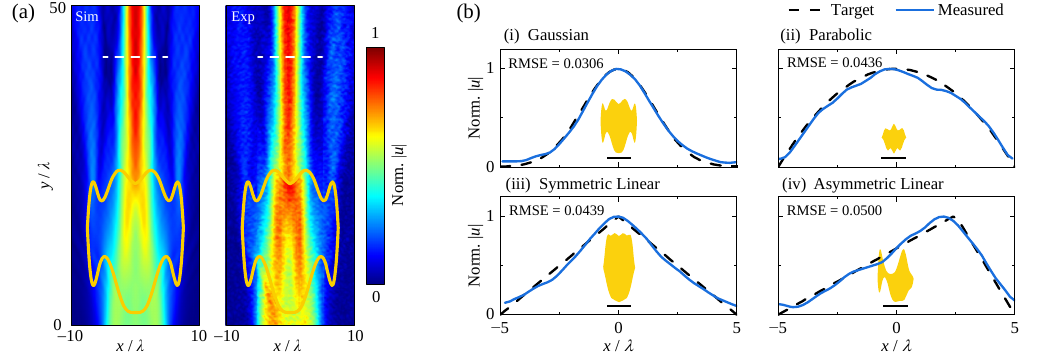}
  \caption{Experimental validation of SAW amplitude shaping. (a)~Simulated and LDV-measured normalized out-of-plane displacement amplitudes for a Gaussian target. The yellow curves outline the optimized CDAL, and the white dashed lines mark the target axial position.
(b)~Target and measured normalized amplitude profiles for different targets. The yellow insets show the optimized CDAL contours. The scale bars in the insets correspond to $10\lambda$, the IDT aperture. }
  \label{fig2}
\end{figure*}

For demonstration, we first designed a CDAL that converts an incident SAW beam into an on-chip focused Gaussian beam, and compared the designed field with the full-field distribution measured by laser Doppler vibrometry~(LDV). As shown in Fig.~\ref{fig2}(a), both simulation and experiment capture the progressive lateral redistribution within the CDAL and the lateral Gaussian distribution of the SAW at the target axial position. Within the metallized region, the measured out-of-plane displacement is slightly enhanced relative to simulation, owing to a weak mismatch between the local and free-surface SAW modes~\cite{yamada1990Scattering}. Despite this discrepancy, the overall diffractive evolution and lateral-field reconstruction agree well with expectations. In this experiment, the SAW beam was excited by a $10\lambda$-aperture straight IDT. For such a beam on unpatterned free surface, comparable diffraction-induced lateral redistribution would normally require several tens of wavelengths of propagation~\cite{zhang2021Field}. The CDAL reshapes this diffractive evolution and establishes the target lateral profile over a shorter axial distance.\par

Furthermore, we designed CDALs that shape SAW beams into parabolic, symmetric-linear, and asymmetric-linear lateral profiles. The same straight IDT served as the source. Fig.~\ref{fig2}(b) compares the target and measured normalized amplitude profiles at the target axial position for all four cases. The measured profiles closely follow the expected distribution, demonstrating that continuous diffractive modulation extends beyond Gaussian focusing to enable precise reconstruction of diverse lateral SAW profiles.\par

\begin{figure}[htbp]
\centering
  \includegraphics[width=8.6cm]{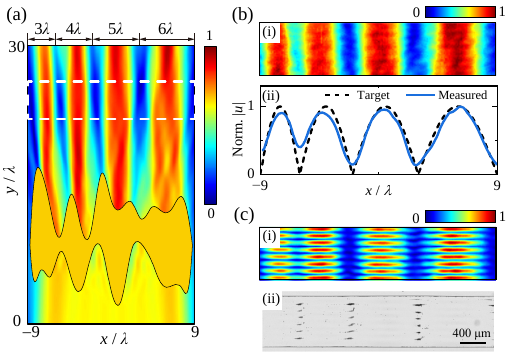}
  \caption{(a) Simulated normalized displacement field of a SAW beam split into four discrete lateral channels. The yellow region denotes the optimized CDAL, and the white dashed rectangle marks the target region detailed in (b) and (c). (b)~LDV-measured normalized amplitude in the target region. (c)~Modulated standing SAW and the corresponding two-dimensional patterning of $5~\mathrm{\upmu m}$ polystyrene particles.}
  \label{fig3}
\end{figure}

We next applied the CDAL to customized beam splitting of a broad incident SAW. The simulation in Fig.~\ref{fig3}(a) shows that an $18\lambda$-aperture SAW beam was split by the CDAL into four adjacent channels with characteristic widths of $3\lambda$, $4\lambda$, $5\lambda$, and $6\lambda$. The target profile was constructed by concatenating four half-period cosine segments of different widths and equal amplitudes. The LDV measurements in Fig.~\ref{fig3}(b) closely reproduce the designed splitting pattern, yielding an RMSE of 0.0962 between the measured and target profiles at the central cross section of target region.\par

Importantly, the splitting preserves the lateral phase distribution well enough for two counter-propagating, CDAL-shaped SAWs to interfere directly. Their superposition forms the two-dimensional, laterally aperiodic standing SAW field shown in Fig.~\ref{fig3}(c)(i). This field is generated using only one pair of acoustic sources, rather than two pairs, and its customizable lateral distribution is difficult to achieve with conventional methods. This linear field distribution further enables unconventional exploitation of nonlinear effects. For example, in the acoustofluidic experiment shown in Fig.~\ref{fig3}(c)(ii), the field produces laterally nonuniform Gor'kov potential wells~\cite{shen2026Sizedependent} that assemble $5~\mathrm{\upmu m}$ polystyrene particles in the microchannel into a laterally non-equidistant pattern, providing a direct mechanical manifestation of the engineered lateral field distribution.\par

\begin{figure}[htbp]
\centering
  \includegraphics[width=8.6cm]{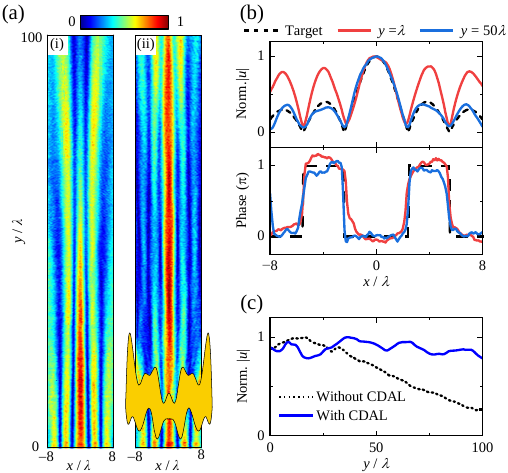}
  \caption{(a)~LDV-measured normalized displacement fields excited by a phase-encoded IDT (i) without and (ii) with CDAL modulation. (b)~Comparison of the ideal targets against the measured amplitude and phase profiles prior to ($y = \lambda$) and after ($y = 50\lambda$) the amplitude reconstruction. (c)~Measured normalized amplitude along the propagation direction at $x = 0$. }
  \label{fig4}
\end{figure}

The CDAL can in principle provide complex-field modulation, but its weak phase accumulation is ill-suited to target fields with abrupt spatial phase variations. Combining it with lateral source-phase encoding provides a more convenient route to such beams. For a zero-order on-chip Bessel beam, for example, generating the discrete $\mathrm{\pi}$-phase jumps between adjacent lobes solely through the CDAL would require pronounced contour variations. We therefore use a lateral phase-encoded IDT~(PE-IDT) to directly encode the discrete $0/\mathrm{\pi}$ phase pattern at the source~\cite{ma2024Generating,li2025Onchip}, while the CDAL reconstructs the corresponding lateral amplitude during propagation.\par

Fig.~\ref{fig4}(a) compares the LDV-measured fields produced by the PE-IDT alone and by the combination of PE-IDT and CDAL. With source-phase encoding alone, the required lateral phase distribution is already established within the aperture, but the amplitude profile departs markedly from the target Bessel function, leading to observable beam spreading during propagation. With the CDAL, continuous diffractive modulation reshapes the lateral amplitude while preserving the encoded phase distribution. As shown in Fig.~\ref{fig4}(b), at $y=\lambda$ the field exhibits the desired step-like phase structure but an incorrect amplitude profile; at $y=50\lambda$, the amplitude approaches the target while the $\mathrm{\pi}$ phase differences between adjacent lobes are retained. Consequently, beam diffraction is strongly suppressed, and the reconstructed complex field exhibits markedly improved propagation stability. Fig.~\ref{fig4}(c) shows the normalized on-axis amplitude at $x=0$. Without the CDAL, this amplitude decays rapidly with propagation distance; with it, the amplitude remains relatively high over approximately $100\lambda$. These results demonstrate that discrete phase encoding and continuous diffractive reconstruction act synergistically to generate high-quality non-diffracting SAW beams.\par

In summary, we demonstrated an on-chip SAW shaping mechanism based on distributed weak perturbations and continuous diffractive evolution. Metallization-induced phase-velocity perturbations are converted by diffraction into substantial lateral field redistribution. This enabled prescribed field profiles and multichannel splitting with unequal widths. Combined with discrete source-phase encoding, continuous diffractive modulation further enables accurate and convenient modulation of the amplitude and phase of complex fields, such as on-chip Bessel beams. The simple planar implementation and nonresonant nature of the CDAL provide a propagation-domain degree of freedom for integrated elastic-wave modulation. More generally, extending the inverse design to constrain spatial phase evolution could further enable wavefront engineering and more general complex-field synthesis.

\bibliographystyle{unsrtnat}
\bibliography{Ref_CDAL}  

\end{document}